\documentclass[12pt]{article}

\usepackage{latexsym}
\usepackage{graphicx}
\usepackage{amssymb,amsmath}

\def \eq#1{Eq.~(\ref{#1})}

\def \fig#1{Fig.~\ref{#1}}
\def \figs#1#2{Figs.~\ref{#1}--\ref{#2}}
\def\slash#1{\setbox0=\hbox{$#1$}#1\hskip-\wd0\dimen0=5pt\advance
       \dimen0 by-\ht0\advance\dimen0 by\dp0\lower0.5\dimen0\hbox
         to\wd0{\hss\sl/\/\hss}}
 
\newcommand\bsm{\bar{B}_s\to\mu^+\mu^-}
\newcommand\bsg{\bar{B}\to X_s\gamma}
\newcommand\bsll{\bar{B}\to X_s l^+l^-}
\newcommand\bbb{\bar{B}_s-B_s}
\newcommand\mgl{m_{\tilde{g}}}
\newcommand\tanb{\tan\beta}
\newcommand\dll{\delta_{LL}}
\newcommand\drr{\delta_{RR}}
\newcommand\dlr{\delta_{LR}}
\newcommand\drl{\delta_{RL}}
\newcommand\mdbare{m_{d}^{(0)}} 
\newcommand\mdphys{m_{d}} 
\newcommand\mbphys{m_{b}} 
\newcommand\delmbs{\Delta M_{B_s}}

\newcommand\EWSym{{\rm SU}(2)_L\times{\rm U}(1)_Y}

\newcommand{\newc}{\newcommand}
\newc\eg{{\it {e.g.}}}	\newc\vs{{\it {vs.}}}	
\newc\etc{{\it {etc.}}}	\newc\ie{{\it {i.e.}}}
\newcommand\tev{\,\mbox{TeV}}
\newcommand\gev{\,\mbox{GeV}}

\newcommand{\msquark}{m_{\widetilde{q}}}
	
\begin{document}

\begin{titlepage}
\pagestyle{empty}
\baselineskip=21pt
\rightline{} 
\vskip 1.5cm
\begin{center}

{\large \bf New Higgs Effects in B--Physics
  in Supersymmetry\\with General Flavour Mixing}

\end{center}   
\begin{center}   
\vskip 0.75 cm
{\bf John Foster}${}^a$, {\bf Ken-ichi Okumura}${}^b$ and {\bf Leszek Roszkowski${}^a$}\\
\vskip 0.1in
\vskip 0.4cm
${}^a$ 
{\it Department of Physics and Astronomy, University of Sheffield, 
  Sheffield, UK} \\
${}^b$
{\it Department of Physics, KAIST, Daejeon, 305-701,Korea}

\vskip 1cm
\abstract{We investigate the effect of general flavour mixing among
squarks on the rare decays $\bsg$, $\bsm$ and $\bbb$ mixing beyond the
leading order in perturbation theory. We include all large
$\tanb$--enhanced corrections whilst also taking into account the
effects of general flavour mixing on the uncorrected quark mass matrix
and $\EWSym$ breaking. For $\bsm$ and $\bbb$ mixing we find that, in
analogy to $\bsg$, there appears a focusing effect which can reduce
the contribution due to the $\drr$ (and the $\dll$) insertion by up to
a factor of two at large $\tanb$ and $\mu>0$.  A dependence on $\dlr$
and $\drl$, that otherwise cancels to first order in the mass
insertion approximation, is also reintroduced.  Taking into account
the current experimental bounds on $\delmbs$ and BR($\bsm$), we find
that the insertions $\drl$ and $\drr$ can be significantly constrained
compared to bounds obtained from $\bsg$ only.}
\end{center}
\baselineskip=18pt \noindent

\vfill
\end{titlepage}

\section{Introduction}

Flavour changing neutral current (FCNC) processes provide a promising
place to look for possible signals of physics beyond the Standard
Model (SM). This is because the GIM mechanism ensures that the SM contributions
and additional effects due to ``new physics'' both enter at the 
one--loop level. As such the increasingly accurate experimental 
data obtained from dedicated $B$--factories as well as collider
experiments can be used to place constraints on masses and other
parameters of a given new physics model.

The process that provides some of the strictest constraints on 
physics beyond the SM is $\bsg$. The current world average for the 
branching ratio is given by~\cite{SLAC:bbm}
\begin{align}
{\rm BR}(\bsg)_{\rm exp}&=(3.52\pm0.30)\times 10^{-4}&
E_{\gamma}>\frac{1}{20}m_b.\nonumber 
\end{align}
The SM prediction for the branching ratio for the decay is based on a
next--to--leading order (NLO) 
calculation that was completed in
Refs.~\cite{GM:bsg,BCMU:bsg}\footnote{A somewhat more conservative  
estimate is given in Ref.~\cite{HLP:bsg}.}
\begin{align}
{\rm BR}(\bsg)_{\rm SM}&=(3.70\pm0.30)\times 10^{-4}&
E_{\gamma}>\frac{1}{20}m_b.\nonumber 
\end{align}

Using the recent results for the decay $\bsll$, the sign of the $\bsg$
amplitude can also be determined~\cite{GHM:bsg} to be that of the SM
contribution. These results allow further constraints to be placed on
any new physics beyond the SM that feature large
contributions from additional sources of flavour violation.

Studies of physics beyond the SM such as supersymmetry (SUSY) have,
until recently, typically focused on the inclusion of leading order
(LO) effects (for example, see Ref.~\cite{LO:bsg}). However, due to
the increasing accuracy of experimental data and its relatively good
agreement with the SM prediction it is becoming necessary to include
at least the dominant effects that occur beyond the LO.

Such effects have been studied, for example, in the two Higgs doublet 
model (2HDM) and the Minimal Supersymmetric Standard Model (MSSM). 
The 2HDM calculation was completed in Refs.~\cite{CDGG1:bsg,BG:bsg}.
Studies of the MSSM contributions have tended to focus on various 
approximations and specific schemes. For instance, the results
presented in Ref.~\cite{CDGG2:bsg}  
assume minimal flavour violation (MFV) and a particle spectrum in 
which the charginos and one of the top squarks are lighter than 
the rest of the superpartners. In Refs.~\cite{DGG:bsg,CGNW:bsg}
the effect of large $\tanb$--enhanced beyond leading order (BLO)
corrections to the $b$--quark
mass and charged Higgs coupling
on the process $\bsg$ was explored and shown to
be sizable, and a resummation of terms proportional to $\alpha_s\tanb$
was performed to keep perturbation expansion under control.
These methods were extended to include neutral Higgs 
contributions and $\tanb$--enhanced corrections to the tree--level 
CKM matrix in Ref.~\cite{AGIS:bdec}, general electroweak 
contributions and $\EWSym$ breaking effects in Ref.~\cite{BCRS:bdec} 
and the effects of general flavour mixing (GFM) in the soft sector 
in Ref.~\cite{OR2:bsg}. 

The calculation presented in Ref.~\cite{OR2:bsg} was based on constructing
an effective field theory where the supersymmetric particles are
integrated out at a scale $\mu_{SUSY}$ (in a similar manner to 
Ref.~\cite{DGG:bsg}). The down quark tree--level (or, in the
language of Ref.~\cite{OR2:bsg}, ``bare'') mass matrix and the effective
couplings were then calculated in the context of GFM. It was found
that taking into account these effects can significantly reduce the 
bounds on the flavour violating parameters compared to purely LO 
calculations as a result of a ``focusing effect''~\cite{OR2:bsg}.

In this Letter we extend and generalise the methods presented
in~\cite{OR2:bsg} to $\bbb$ mixing and the decay $\bsm$. 
Whilst the mass difference $\delmbs$ and the branching ratio
BR($\bsm$) have so far remained undetermined, both processes provide
possible places to look for signals of physics beyond the SM.
In particular, large regions of the  MSSM in the regime of large
$\tanb$ can be explored. Since $B$--factories do  
not run at the  energy required to produce large quantities of the 
$B_s$ mesons, the best experimental constraints on both processes are 
provided by collider experiments. The current experimental bound for the 
process $\bsm$ is provided by the D\O~group at the Tevatron~\cite{D0:bsm}
\begin{align}
{\rm BR}(\bsm)_{\rm exp}&<5.0\times 10^{-7}& 95\% {\rm~C.L},\nonumber
\end{align}
whilst the experimental bound on $\delmbs$ is~\cite{SLAC:bbm}
\begin{align}
\Delta M_{B_s}^{\rm exp}&>14.5{\rm ps}^{-1}& 95\% {\rm~C.L}.\nonumber
\end{align}
The SM contributions to these processes are known to NLO
\cite{NLO:bsm,NLO:bbm}. However, the largest source of error
for both quantities is due to the hadronic matrix element
$f_{B_s}$ which has to be determined using either lattice calculations
or QCD sum rules. The values given in the literature for the branching 
ratio for the process $\bsm$ therefore tend to vary but are typically 
of the order~\cite{Buras:rev}
\begin{align}
{\rm BR}(\bsm)_{\rm SM}=(3.2\pm1.5)\times10^{-9}.
\end{align}
The SM prediction for $\delmbs$ is~\cite{BCJL:bbm}
\begin{align}
\delmbs^{\rm SM}=18.0\pm3.7{\rm ps}^{-1},
\end{align}
where the large hadronic uncertainty has been avoided by using the
experimentally measured value of $\Delta M_{d}^{\rm SM}$. It has also
been pointed out that in models with minimal flavour violation
the large uncertainty associated with the branching ratio for
the decays $\bar{B}_q\to\mu^+\mu^-$ can also be avoided by relating
it to the experimentally measured values of $\Delta M_{B_q}$
\cite{Buras:bsm}.

The contributions due to effects beyond the SM on the decay $\bsm$
arise due to contributions to $Z$ and Higgs penguins as well as box
diagrams mediated by charginos and (in the GFM framework)
neutralinos. The contributions due to neutral Higgs penguins in
particular have been the subject of intense theoretical
investigation~\cite{BK:bsm,MFV:bsm,IR1:bsm,GFM:bsm,IR2:bsm,DP:2002er,AGIS:bdec,BCRS:bdec} 
recently due to the $\tan^6\beta$ dependence of ${\rm BR}(\bsm)$.  At
large $\tanb$ it is therefore quite possible for ${\rm BR}(\bsm)$ to
be enhanced by a few orders of magnitude compared to the SM value,
whilst still satisfying current experimental bounds and the
restrictions imposed by $\bsg$. Similar contributions arise in the
$\bbb$ system, where the double Higgs penguin diagram, although
strictly an NLO effect, can become comparable to the LO contributions
in the large $\tanb$ limit~\cite{BCRS:bbm,IR2:bsm,BCRS:bdec}.

In this Letter, 
in addition to the effects discussed in~\cite{OR2:bsg},
we include the contributions of charginos and neutralinos when
calculating corrections to the bare quark mass matrix and corrected
vertices. We also include the effects of the modified neutral
Higgs vertex when evaluating the contributions to $\bsg$. For
all three processes: $\bbb$ mixing and the decays $\bsg$ and $\bsm$,
we take into account all $\tanb$--enhanced 
corrections, the additional electroweak and $\EWSym$ breaking
effects discussed in~\cite{BCRS:bdec} and the effects of GFM
mixing parameters on the bare mass matrix~\cite{OR2:bsg}. 

\section{GFM and {\boldmath $\tanb$}--Enhanced Corrections}

The effects of $\tanb$--enhanced SUSY corrections to the down quark Yukawa
coupling~\cite{HRS:qme} and the charged Higgs coupling~\cite{DGG:bsg,CGNW:bsg}
have been found to be large and their inclusion can produce sizable
deviations from purely LO calculations. As has been pointed out
in Refs.~\cite{BK:bsm,IR1:bsm,AGIS:bdec,BCRS:bdec}, these corrections can 
also affect the structure of the CKM matrix, K, due to the additional 
unitary transformations required to transform the quark fields into 
a super--CKM basis.

To illustrate this consider the effect of integrating out the SUSY
particles on the down quark mass matrix. The physical down quark masses
(denoted $\mdphys$) are given in terms of the uncorrected quark masses 
(denoted $\mdbare$) by the relation\footnote{We will generally follow
  the language and conventions of Ref.~\cite{OR2:bsg}.} 
\begin{equation}
\mdphys=\mdbare+\delta m_d.\label{qmc:eq2}
\end{equation}
Note that, relative to Ref.~\cite{OR2:bsg}, we have dropped the factor
$\alpha_s/4\pi$ in front of $\delta m_d$ because here we will also
include chargino and neutralino corrections, in addition to the
(dominant) SUSY QCD ones considered in Ref.~\cite{OR2:bsg}.

In the physical super--CKM basis the mass matrix $\mdphys$ is (by
definition) diagonal, but in general $\mdbare$ (and $\delta m_d$) is
not and provides a source of flavour violation. 
Alternatively~\cite{BK:bsm,IR1:bsm,AGIS:bdec,BCRS:bdec}, one can
start with the ``bare'' super--CKM basis where $m_d^{(0)}$ is diagonal
and, after computing the corrections, diagonalise the corrected mass
matrix, which amounts to rotating to the physical super--CKM
basis. In this approach flavour violation is introduced via unitary
rotation matrices as they affect the CKM matrix, as well as the
various other vertices present in the resulting effective theory.

In the limit of MFV both approaches can be shown to be equivalent. For
example, the $1+\epsilon_s\tanb$ dependence (where $\epsilon_s$ will
be defined shortly) that arises when the CKM matrix is modified in the
approach described in Refs.~\cite{BK:bsm,IR1:bsm,AGIS:bdec,BCRS:bdec}
is reproduced once the gluino contributions to a given process are
taken into account. When performing MFV calculations the method
presented in Ref.~\cite{BCRS:bdec} is more convenient since the gluino
corrections to a given vertex are solely to the flavour diagonal
terms. However, in GFM scenarios the flavour violating gluino (and
neutralino) contributions are evaluated anyway. Additionally the
iterative procedure described in Ref.~\cite{OR2:bsg} is  more
suited to calculations where the squark mass matrix is diagonalised
numerically.

Here we follow the procedure described in Ref.~\cite{OR2:bsg}. The
bare mass matrix $\mdbare$ and the corrections to the electroweak
vertices are calculated in the physical super--CKM basis using an
iterative procedure. The supersymmetric contributions to the process
in question are then evaluated, taking into account the effects of the
modified bare mass matrix, and evolved from $\mu_{SUSY}$ to the
electroweak scale $\mu_{W}$ using the relevant six flavour anomalous
dimension matrix. The electroweak contributions are then evaluated,
using the uncorrected vertices when evaluating the NLO corrections and
the corrected vertices when evaluating the LO contributions. Finally
the combined supersymmetric and electroweak Wilson coefficients are
evolved from $\mu_{W}$ to $\mu_b$ using the five flavour anomalous
dimension matrix and used to calculate the relevant observable for the
process in question.

Before presenting our numerical results it will be useful to consider
the effects of including GFM contributions when calculating $\mdbare$
and the resulting effects on the Wilson coefficients relevant to
$\bsg$, $\bsm$ or $\bbb$ mixing. To this end we work in the mass
insertion approximation (MIA) where the off--diagonal entries of the
$6\times 6$ squark mass matrix are treated as perturbations and
flavour violation is communicated through mixed propagators
proportional to the appropriate off--diagonal element.  In our
numerical analysis MIA is not assumed and all the squark mass matrices
are diagonalised numerically.

Departures from MFV can be measured in terms of the parameters $\dll$, $\dlr$,
$\drl$ and $\dlr$ definitions of which can be found in Ref.~\cite{OR2:bsg}.
If one delta is varied at a time the diagonal terms of the bare mass matrix
are given by the well known result~\cite{HRS:qme}
\begin{align}
\left(\mdbare\right)_{ii}=
\frac{\left(\mdphys\right)_{ii}}
{1 + \left(\epsilon_s+\delta_{i3}\epsilon_Y Y_t^2\right)\tanb},\label{mdbare:diag} 
\end{align}
where $i=1,2,3$ (or $d$, $s$, $b$), $Y_t$ is the Yukawa coupling of the top
quark and the
presence of the Kronecker $\delta$--function $\delta_{i3}$ reflects
the fact that the chargino contribution proportional to the top quark
Yukawa coupling only effects the bottom quark mass. (Corrections to
the strange and down quark masses are suppressed by $K_{ts} K_{ts}^\ast$ and
$K_{td} K_{td}^\ast$, respectively, and are set equal to zero.)
Finally, 
\begin{align}
\epsilon_s&=-\frac{\alpha_s}{2\pi} C_2(3)
\frac{\mu}{\mgl} 
H_2(x_{\tilde{d}_R},x_{\tilde{d}_L}),
&\epsilon_Y&=-\frac{A_t}{16\pi^2\mu}
H_2(y_{\tilde{u}_R},y_{\tilde{u}_L}),
\end{align}
where $\alpha_s$ is the strong coupling constant,
$C_2(3)= 4/3$ is the quadratic Casimir operator for $SU(3)$,
$\mu$ is the Higgs/higgsino mass parameter, $\mgl$ is the gluino
mass and $A_t$ is the $3,3$ element
of the trilinear up--type soft term. The loop function $H_2$
can be found in the Appendix. Its arguments and some other quantities
to appear below are defined as
\begin{align}
x_{\tilde{d}_{L,R}}&=\frac{\widetilde{m}_{d,LL,RR}^2}{\mgl^2},
&
x_{\tilde{d}_{RL}}& =
\frac{\sqrt{\widetilde{m}_{d,LL}^2}\sqrt{\widetilde{m}_{d,RR}^2}}{\mgl^2},
& y_{\tilde{u}_{L,R}}& = \frac{\widetilde{m}_{u,LL,RR}^2}{\mu^2},
\end{align}
where $\widetilde{m}^2_{d,LL}$, $\widetilde{m}^2_{d,RR}$ denote
common values of the diagonal entries of $3\times3$ squark
soft SUSY breaking terms for which we follow the conventions given in
Ref.~\cite{OR2:bsg}. Whilst the diagonal elements of the soft terms
have been assumed to be universal, it is relatively easy to include
the effects of flavour dependence at the cost of clarity in the final
expressions.

It has been pointed out in Ref.~\cite{Demir:bsm} that
if $\dll$ and $\drr$ are both non--zero large corrections to the bare
strange and down quark masses can occur at third order in the MIA. 
In our numerical analysis we diagonalise the squark mass matrix
numerically and therefore these corrections are automatically
included.

Taking into account flavour violating effects in the LR sector and 
ignoring the effects induced by other sources of flavour violation 
(including the CKM matrix) the off--diagonal elements of $\mdbare$ are given 
by (a more complete formula will be given in Ref.~\cite{for2:bdec})
\begin{align}
\left(\mdbare\right)_{ij}=
\frac{\epsilon_{RL}}{1+\left(\epsilon_s+\epsilon_Y
  Y_t^2\delta_{j3}\right)\tanb}\;x_{\tilde{d}_{RL}}\mgl
\left(\drl\right)_{ij} +  \mathcal{O}(\delta^3)\label{mdbare:offdiag}
\end{align}
where $i,j=1,2,3$ and 
\begin{equation}
\epsilon_{RL}= -\frac{\alpha_s}{2\pi} C_2(3)
H_2(x_{\tilde{d}_R},x_{\tilde{d}_L}).
\end{equation}
The effect of including GFM corrections to
$\mdbare$ and the electroweak vertices can be rather large. For example,
in the case of $\bsg$ the presence of non--zero
$\left(\mdbare\right)_{32}$ can 
lead to large corrections to $\delta^{\chi^-} C_{7,8}$ that are 
otherwise suppressed by a factor of $K_{cb}m_b$~\cite{OR2:bsg}.
Similarly $\left(\mdbare\right)_{23}$ induces analogous corrections to the 
chargino contributions in the primed sector that are usually
suppressed by $m_s K_{ts}^\ast$.

In the case of $\bsm$, the gluino contributions to the Wilson 
coefficients of the scalar and pseudoscalar operators become
(in the large $\tanb$ limit)
\begin{align}
\delta^{\tilde{g}}C_{S,P}=\pm\frac{4\alpha_s}{3\alpha}\frac{\mu
  m_{\mu}} {m_A^2 {m}_b\mgl} \frac{\tan^3\beta}{K_{tb}K_{ts}^{\ast}}
  \bigg[&\left(\mdbare\right)_{32}^{\ast} 
    H_2(x_{\tilde{d}_R},x_{\tilde{d}_L})\nonumber\\
    +&\left(\mdbare\right)_{33}^{\ast}\left(\dll\right)_{23}
    x_{\tilde{d}_L}
    H_3(x_{\tilde{d}_R},x_{\tilde{d}_L},x_{\tilde{d}_L})\bigg],
\label{CSP:bsm}  
\end{align}
where $m_{\mu}$ denotes the mass of the $\mu$--lepton, 
whilst the primed coefficients are given by
\begin{align}
\delta^{\tilde{g}}C_{S,P}^{\prime} = \frac{4\alpha_s}{3\alpha}
\frac{\mu m_{\mu}}{m_A^2 {m}_s\mgl}
\frac{\tan^3\beta}{K_{tb}K_{ts}^{\ast}}
\bigg[&\left(\mdbare\right)_{23} H_2(x_{\tilde{d}_L},x_{\tilde{d}_R})\nonumber\\
+&\left(\mdbare\right)_{33}\left(\drr\right)_{23} x_{\tilde{d}_R} 
H_3(x_{\tilde{d}_L},x_{\tilde{d}_R},x_{\tilde{d}_R})\bigg],
\label{CSPpr:bsm}
\end{align}
(for details of the operator basis we use see Ref.~\cite{BEKU2:bsm}).
Substituting \eq{mdbare:diag} and \eq{mdbare:offdiag} into the above
expressions yields the Wilson coefficients:

\begin{align}
\delta^{\tilde{g}}C_{S,P}=&\pm\frac{4\alpha_s}{3\alpha}\frac{\mu
  m_{\mu}} {m_A^2} \frac{\tan^3\beta}{K_{tb}K_{ts}^{\ast}}
  \Bigg\{
\frac{\epsilon_{RL}}{[1+\epsilon_s\tanb]}
\frac{x_{\tilde{d}_{RL}}}{m_b}
\left(\dlr\right)_{23}H_2(x_{\tilde{d}_R},x_{\tilde{d}_L})\nonumber\\
+&\frac{1}{[1+\epsilon_s\tanb][1+\left(\epsilon_s+\epsilon_Y Y_t^2\right)\tanb]}
\frac{x_{\tilde{d}_L}}{\mgl}
\left(\dll\right)_{23} H_3(x_{\tilde{d}_R},x_{\tilde{d}_L},x_{\tilde{d}_L})\Bigg\},
\label{CSP:bsm2}  
\end{align}
\begin{align}
\delta^{\tilde{g}}C_{S,P}^{\prime} =& \frac{4\alpha_s}{3\alpha}
\frac{\mu m_{\mu}}{m_A^2}
\frac{\tan^3\beta}{K_{tb}K_{ts}^{\ast}}
\Bigg\{\frac{[\epsilon_{RL}+\frac{\mgl}{\mu}\epsilon_Y
    Y_t^2]}{[1+\left(\epsilon_s+\epsilon_Y Y_t^2\right)\tanb]} 
\frac{x_{\tilde{d}_{RL}}}{m_s} 
\left(\drl\right)_{23}H_2(x_{\tilde{d}_L},x_{\tilde{d}_R})\nonumber\\
&+\frac{1}{[1+\left(\epsilon_s+\epsilon_Y Y_t^2\right)\tanb]^2}
\frac{x_{\tilde{d}_R}}{\mgl}\frac{m_b}{m_s}
\left(\drr\right)_{23}H_3(x_{\tilde{d}_L},x_{\tilde{d}_R},x_{\tilde{d}_R})\Bigg\},
\label{CSPpr:bsm2}
\end{align}
where $m_A$ denotes the mass of the pseudoscalar Higgs boson and $\alpha$
denotes the electromagnetic coupling constant.

Comparing the above expressions with the analysis of
Ref.~\cite{IR2:bsm}, the term in Eq. (4.15) proportional to the
insertion\footnote{Since we are interested here in the $23$ mixing
elements, henceforth we shall adopt the conventional notation of
$\left(\dlr\right)_{23}$ as $\dlr$, $\left(\drl\right)_{23}$ as
$\drl$, \etc} $\dll$ needs to be corrected by a factor
\begin{equation}
\frac{1+\left(\epsilon_s+\epsilon_Y Y_t^2\right)\tanb}{1+\epsilon_s\tanb}
\end{equation}
which reflects the additional contribution (which was omitted in
Ref.~\cite{IR2:bsm}) obtained when including the effects of the
insertion on the bare CKM matrix.

Values of up to $\mathcal{O}(1)$ for both $\dll$ and $\drr$ are viable
in some regions of parameter space~\cite{CFMS:bdec} and can lead to
large contributions to $C_S^{(\prime)}$ and $C_P^{(\prime)}$.  Large
values of $\drr$, in particular, are motivated by $SO(10)$ or $SU(5)$
based solutions~\cite{CMM:neut,drr:other} to the neutrino mass
problem.  As well as these corrections, the insertions $\dlr$ and
$\drl$ reappear once BLO effects are taken into account. At LO the
insertions vanish due to an accidental cancellation between the self
energy and vertex corrections to the effective Higgs vertex. Including
BLO effects however, allows the insertions to appear through their
effects on the bare mass matrix \eq{mdbare:offdiag}.  This effect is
independent of the LO cancellation and originates from the additional
unitary transformations that are required to transform between the
bare super--CKM basis and the physical super--CKM basis.
Additionally, since the corrections proportional to $\dlr$ and $\drl$
depend on $\mgl$, rather than the strange or bottom quark mass, the
enhancement by $\mgl/\mbphys$ can compensate for the
$\alpha_s^2$--dependence of the Wilson coefficient.

As discussed in Ref.~\cite{BCRS:bdec} the corrected neutral Higgs vertex
can effect $\bbb$ mixing via double Higgs penguin diagrams that
contribute to the Wilson coefficients of the operators
\begin{align}
\mathcal{O}^{LR}_2=&(b^{\alpha}P_L s^{\alpha})(b^{\beta}P_R s^{\beta}),
&\mathcal{O}^{SLL}_1=&(b^{\alpha}P_L s^{\alpha})(b^{\beta}P_L s^{\beta}),
&\mathcal{O}^{SRR}_1=&(b^{\alpha}P_R s^{\alpha})(b^{\beta}P_R s^{\beta}).
\end{align}
(for details of the operator basis we use see Ref.~\cite{BJU:bbm}).
In the large $\tanb$ regime the dominant contribution arises from the Wilson
coefficient $C^{LR}_2$, as the pseudoscalar and scalar contributions
approximately cancel for both $C^{SLL}_1$ and $C^{SRR}_1$.
For non--zero $\dll$ and $\dlr$ the contributions to $C^{LR}_2$ are
typically suppressed by a factor of the strange quark mass. For 
the insertions $\drr$ and $\drl$, however, it is possible to avoid this
suppression factor via the diagram where one Higgs penguin is mediated
by chargino exchange and the other by gluino exchange. In the case of
non--zero $\drr$, for example, $C^{LR}_2$ is given by
\begin{align}
C^{LR}_2= -\frac{8\alpha_s}{3\pi^2\alpha\sin^2\theta_W}
\frac{A_t Y_t^2 \mbphys^2}{\mgl m_A^2} 
\frac{\tan^4\beta}{K_{tb}^\ast K_{ts}}\;
\kappa\,\left(\drr\right)_{23} \, x_{\tilde{d}_R} 
H_2(y_{\tilde{u}_R},y_{\tilde{u}_L})
H_3(x_{\tilde{d}_L},x_{\tilde{d}_R},x_{\tilde{d}_R}),
\label{dmb:drr}  
\end{align}
where $\kappa$ contains the effects of resuming $\tanb$ enhanced
contributions
\begin{align}
\kappa=\frac{1}{(1+\epsilon_s\tanb) 
\times [1+(\epsilon_s+\epsilon_Y Y_t^2)\tanb]^3}.
\end{align}
\begin{figure}[!tbh]
  \begin{center}
  \begin{tabular}{c c}
    \includegraphics[angle=270,width=0.45\textwidth]{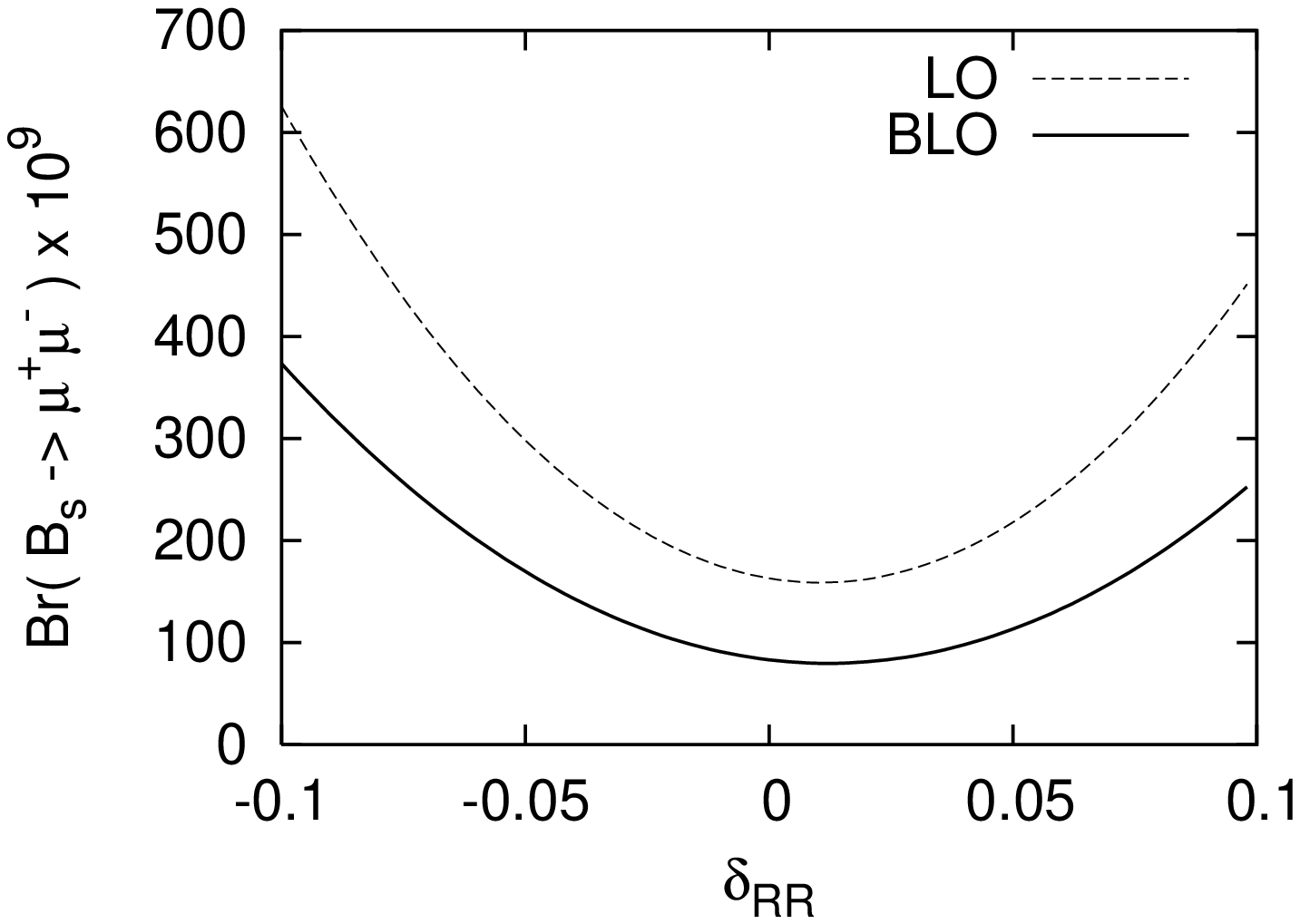}
    & \includegraphics[angle=270,width=0.45\textwidth]{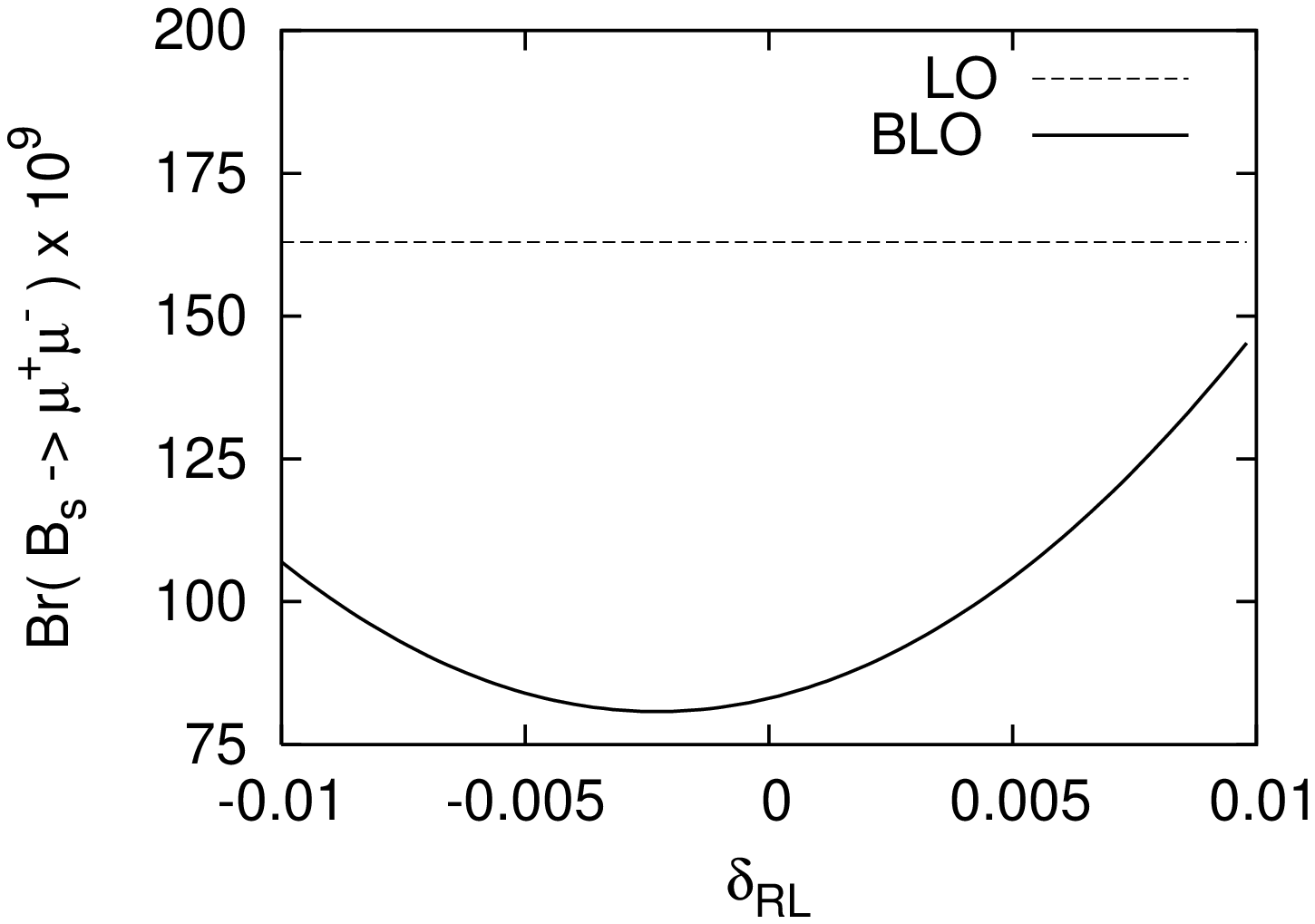}
  \end{tabular}
  \end{center}
  \caption{BR($\bsm$) vs. $\drr$ (on the left) and vs. $\drl$ (on the
  right).The soft sector is parameterised in terms of a common mass for the
   squark soft terms $m_{\tilde{q}}$,
  $\mgl=\sqrt{2}\msquark=1\tev$, $A_t=-1\tev$, $m_{H^+}=\mu=500\gev$ 
  and the gaugino soft terms $M_1=M_2=0.5\mgl$, for $\tanb=50$.}  
  \label{bsm:drrdlrdep}
\end{figure}
%

\section{Numerical Results}

When performing our numerical calculations we diagonalise all the
various mixing matrices numerically, whilst we use {\sl FeynHiggs
2.0.2}~\cite{FeynHiggs} to compute the parameters associated with the
Higgs sector.\footnote {We therefore ignore any possible effects on the
Higgs sector due to off--diagonal entries in the squark mass matrix. 
However, the
corrections due to $\dll$ to the lightest Higgs mass tend to be rather
small~\cite{HHMP:pew}.}  \fig{bsm:drrdlrdep} shows the effects of
including beyond leading order GFM contributions on the decay $\bsm$
and its dependence on the flavour violating parameters
$\drr$ and $\drl$. For $\drr$ the 
effect of using the bare bottom quark mass, as well as the additional
effects induced by the off--diagonal elements of $\mdbare$ present in
\eq{CSPpr:bsm}, reduce the contribution to the decay by about a
factor of two. A similar reduction between the LO and BLO results
occurs for the insertion $\dll$.  This reduction of LO effects by the inclusion
of BLO supersymmetric corrections can be viewed as an extension of
the focusing effect found in Ref.~\cite{OR2:bsg} to the neutral Higgs vertex.
For $\drl$ (and similarly $\dlr$) the effects are more
dramatic. As stated above, at leading order the
contribution due to $\drl$ cancels and the remaining contributions
stem from Z penguin diagrams that are not enhanced at large
$\tanb$. However, once BLO corrections are taken into account the
effects can differ quite significantly from the LO scenario where the
dominant contributions at large $\tanb$ arise from the chargino
contributions to the neutral Higgs vertex.

\begin{figure}[!tbh]
  \begin{center}
  \begin{tabular}{c c}
    \includegraphics[angle=270,width=0.45\textwidth]{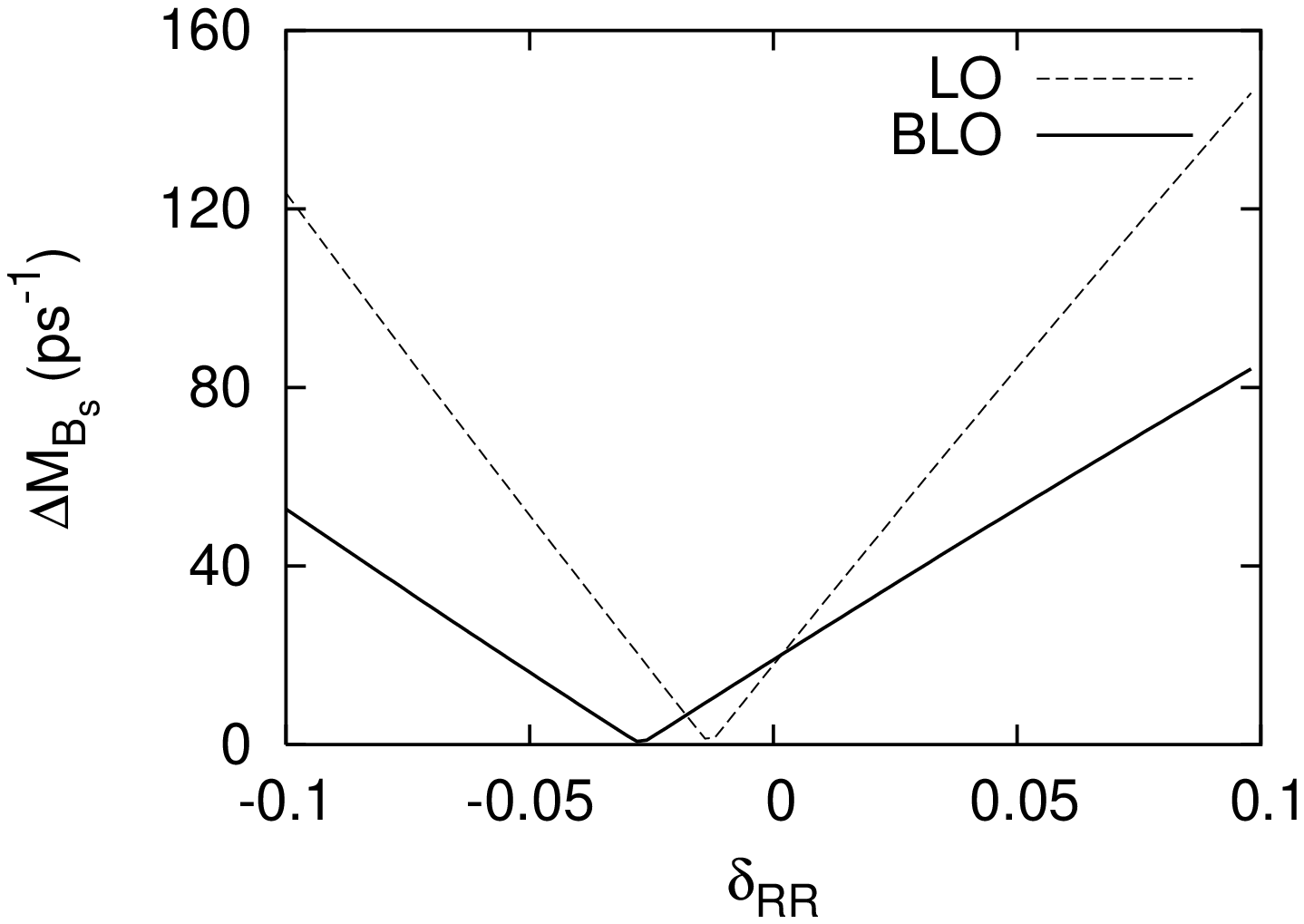}
    & \includegraphics[angle=270,width=0.45\textwidth]{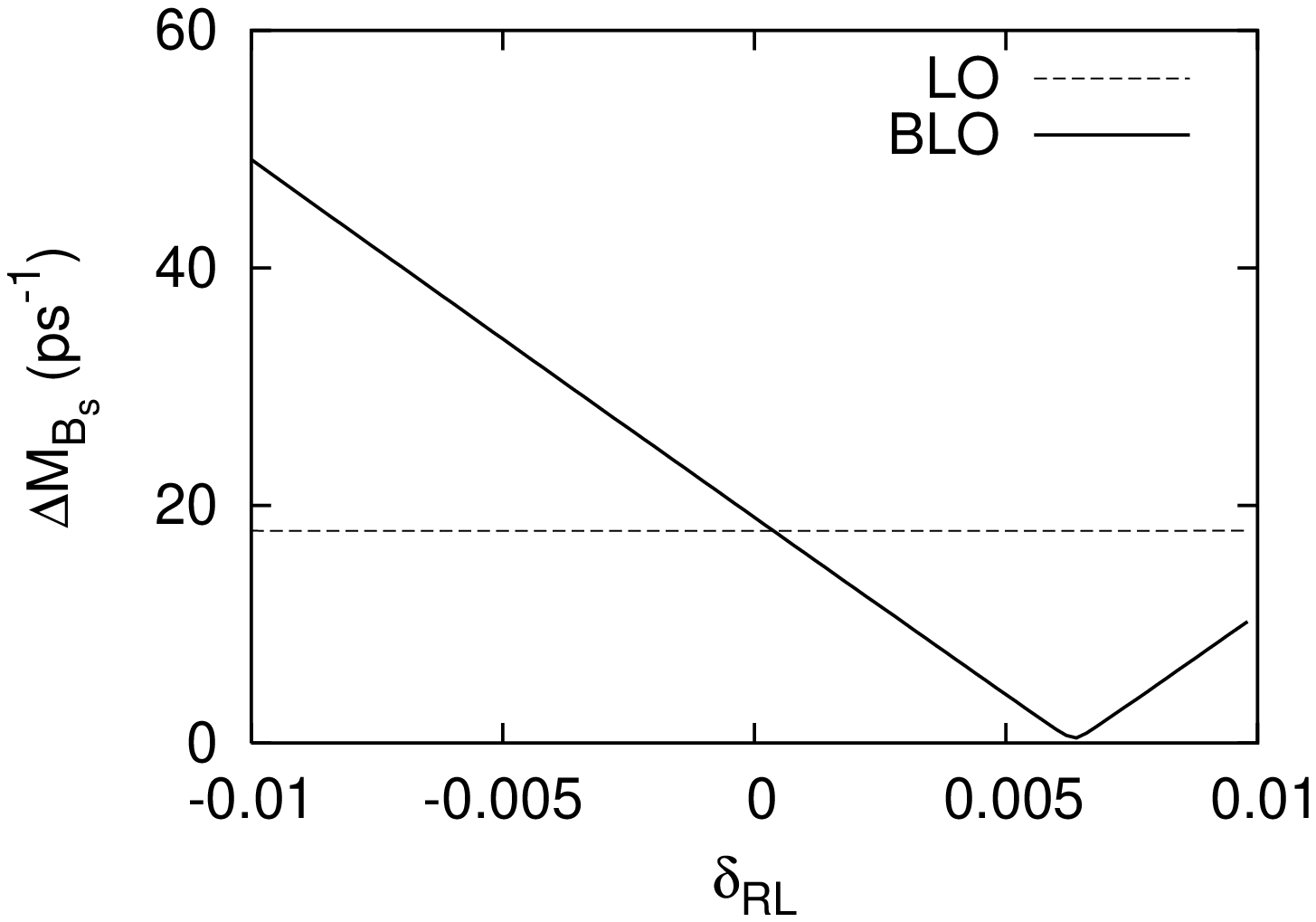}
  \end{tabular}
  \end{center}
  \caption{$\delmbs$ vs. $\drr$ (on the left) and vs. $\drl$ (on the
  right) for the same parameters as in \fig{bsm:drrdlrdep}.} 
  \label{dmb:drrdlrdep}
\end{figure}

\fig{dmb:drrdlrdep} shows a similar plot for the $\bbb$ system. The strong
linear dependence on both $\drl$ and $\drr$ confirms the approximate
formula presented in the previous section. The effect of including BLO
contributions once again has a large effect. Both graphs
are somewhat similar to their $\bsm$ analogs underlining the dependence on the 
corrected Higgs vertex and the large effects it can display in the large
$\tanb$ region. For $\drr$ the effect of including BLO contributions
can once again lessen the dependence on $\drr$ with respect
to a purely LO analysis and lead to values of $\delmbs$ closer to the
SM prediction. In the case of the insertion $\drl$, the effect of 
including BLO contributions is once again rather large and can lead
to large deviations from a purely LO calculation.

\begin{figure}[!tbh]
  \begin{center}
    \includegraphics[width=0.45\textwidth]{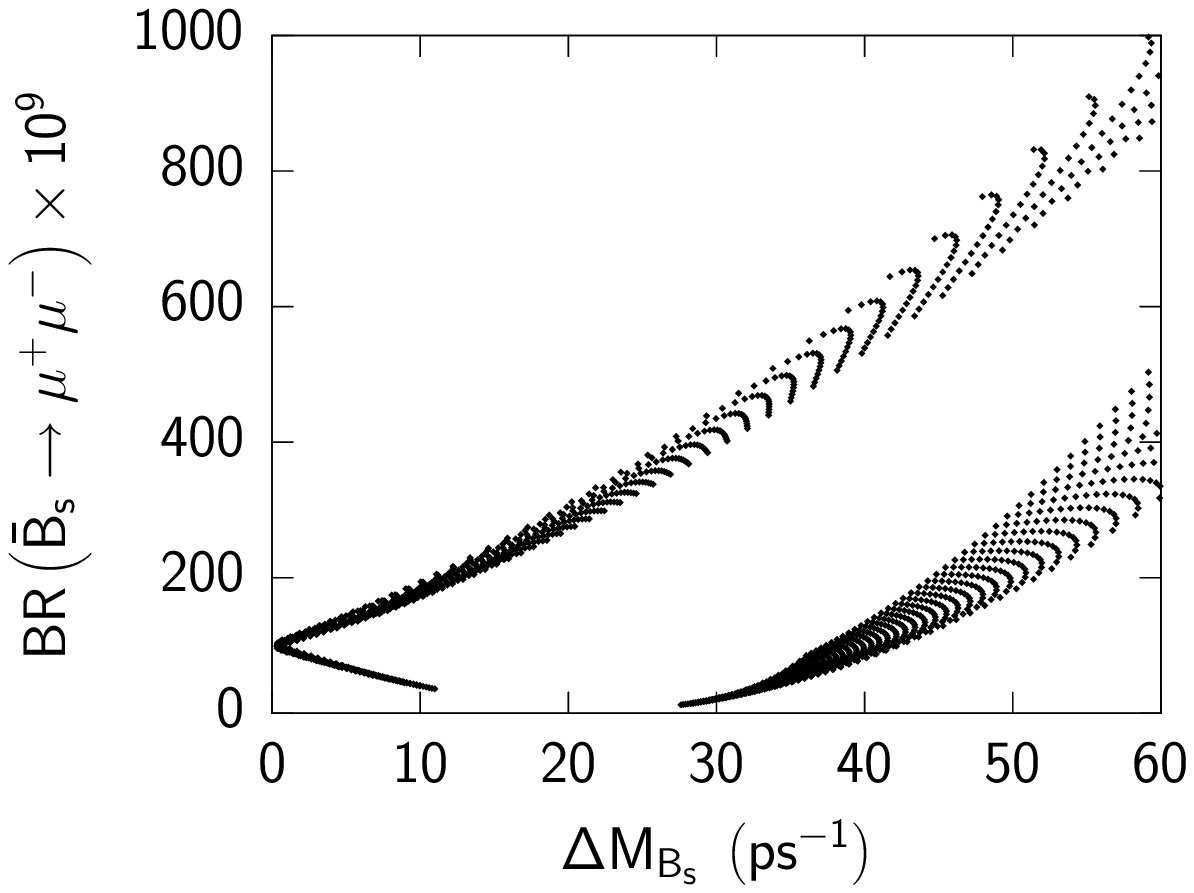}
  \end{center}
  \caption{A scatter plot of BR($\bsm$) vs. $\delmbs$ for
  $0.2<|\drr|<0.5$ and $\tanb=40$.} 
  \label{bsmvsdmb}
\end{figure}

From \figs{bsm:drrdlrdep}{dmb:drrdlrdep} it is evident that the inclusion
of BLO effects can significantly reduce both BR($\bsm$) and $\delmbs$
relative to LO predictions in the case of the insertion $\drr$. This
focusing effect can be viewed as an extension of
the results presented in Ref.~\cite{OR2:bsg} to the neutral Higgs vertex.
\fig{bsmvsdmb} illustrates the strong correlation between $\delmbs$ and
BR($\bsm$) at large $\tanb$ reflecting the fact that both processes are
highly dependent on the neutral Higgs coupling in this regime.
\begin{figure}[!tbh]
  \begin{center}
  \begin{tabular}{c c}
    \includegraphics[width=0.45\textwidth]{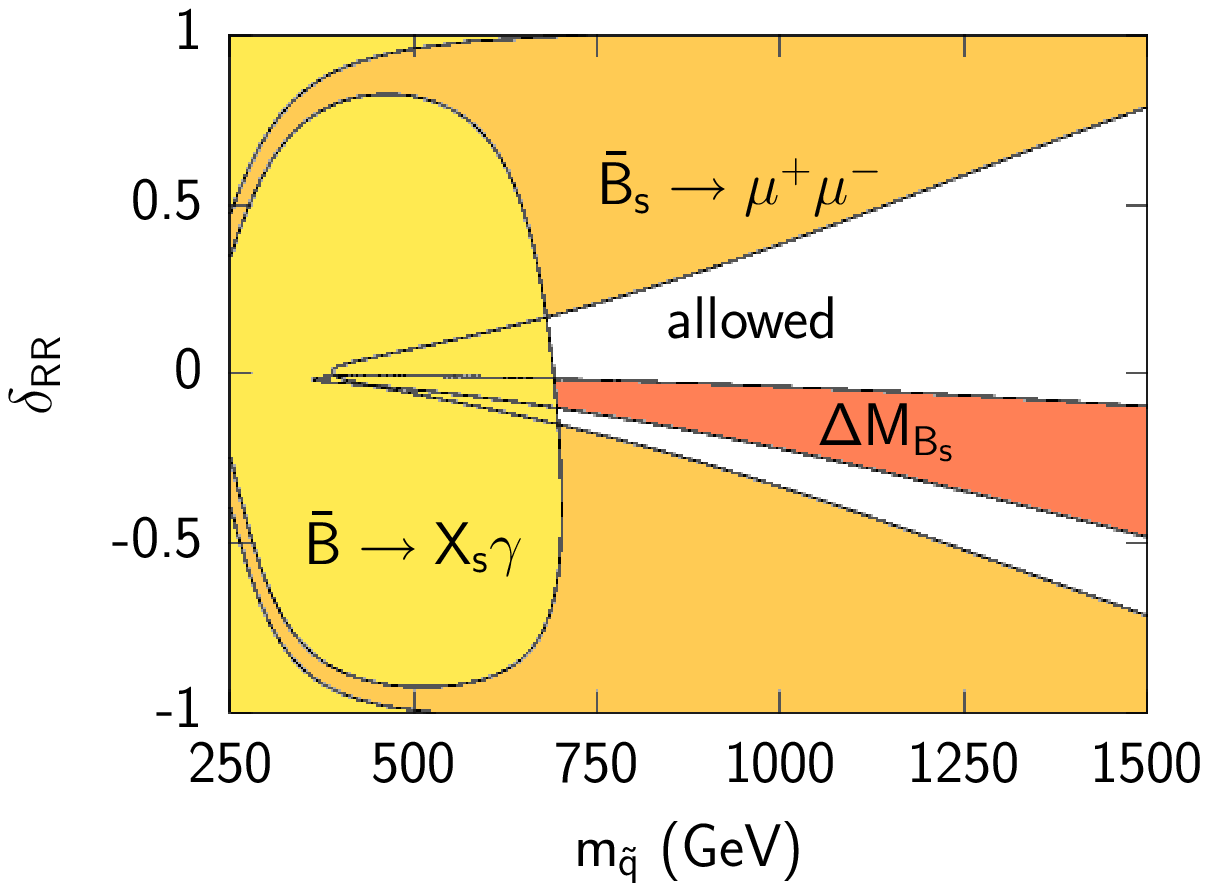}
    & \includegraphics[width=0.45\textwidth]{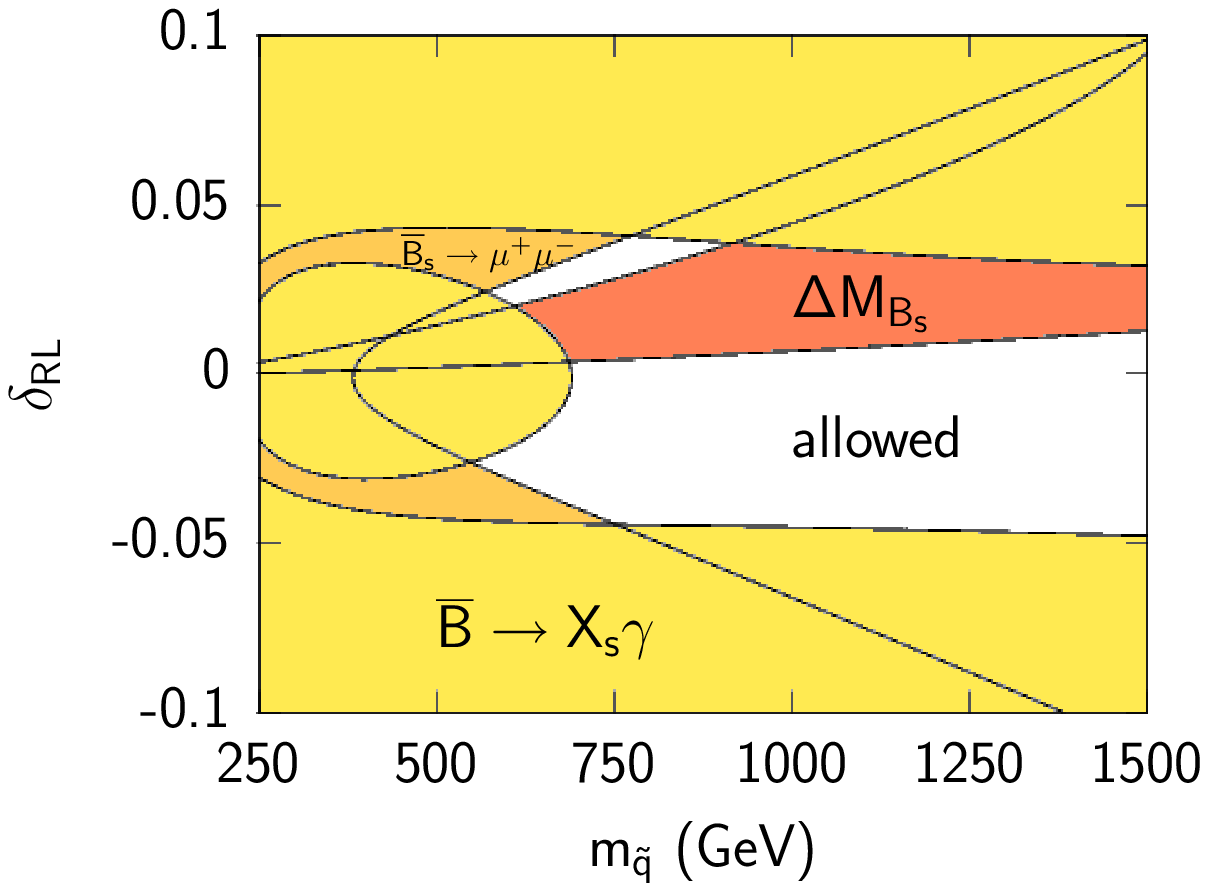}
  \end{tabular}
  \end{center}
  \caption{In the above plots the region excluded by $\bsg$ is shaded in 
    yellow (light grey). The additional regions excluded by
    $\bsm$ and $\delmbs$ are shaded in orange (medium grey) and red
    (dark grey), respectively.
    The soft sector is parameterised as follows,
    $m_{\tilde{g}}=\sqrt{2}m_{\tilde{q}}$,
    $A_t=-m_{\tilde{q}}$ and $\mu=m_{H^+}=0.5m_{\tilde{q}}$, for
    $\tanb=40$.}
  \label{msqvsdxy}
\end{figure}
Finally let us consider the effects the current experimental bounds on
$\bsm$ and $\delmbs$ have on the flavour violating parameters. As discussed
in Ref.~\cite{OR2:bsg} the constraints placed on $\drr$ and $\drl$ by
$\bsg$ are rather weak. This is mainly due to the fact that the dominant
contributions from both insertions are to the primed Wilson coefficients
${\rm C}_{7,8}^{\prime}$. Since there is no interference between the
primed and unprimed operators the contributions to the final branching
ratio are always positive. These additional contributions can therefore,
in the CMSSM (mSUGRA) favoured scenario $A_t<0$, $\mu>0$, counter the
chargino contribution which tends to decrease the overall branching ratio.
These arguments however do not apply to the $\bbb$ system and the decay
$\bsm$, where it is quite possible for the behaviour at large $\tanb$
to be completely dominated by the effects of the Higgs penguin 
contributions.

In \fig{msqvsdxy} we show how the additional constraints supplied by the
decay $\bsm$ and $\delmbs$ affect the otherwise permitted values of
both $\drl$ and
$\drr$. In both plots we used the current 95\% confidence limits
on $\delmbs$ and $\bsm$,
\begin{align}
{\rm BR}(\bsm)&<5.0\times 10^{-7},
&\delmbs>14.5{\rm ps}^{-1}.
\end{align}
For $\bsg$ we combined the current experimental and theoretical errors in
quadrature and added a small additional error to account for the accuracy
of the supersymmetric portion of the calculation
\begin{align}
2.72\times 10^{-4}<{\rm BR}(\bsg)<3.96\times 10^{-4}.
\end{align}
It can be seen from both plots that $\bsm$ and
$\delmbs$ can provide sizable constraints compared to $\bsg$.
$\delmbs$ in particular provides a rather effective
means of constraining $\drl$ and $\drr$. As stated in the previous section,
the contributions due to $\drl$ and $\drr$ to the Wilson coefficient
$C^{LR}_{2}$ are not suppressed by factors of $m_s$. Coupled
with the $\tan^4\beta$ dependence of the contribution the effects of
the double penguin can be rather large and can compete with 
the Standard Model contribution at large $\tanb$.

\section{Conclusions}
We have found that by taking into account GFM contributions when
calculating the radiative corrections to the down quark mass matrix,
the $\dlr$ (and $\drl$) dependence of the corrected neutral Higgs
vertex that conventionally cancels in LO calculations can
reappear. The behaviour of processes that are highly dependent on this
vertex (such as $\bsm$ and $\bbb$ mixing) can therefore change
dramatically once GFM corrections are taken into account.  In the case
of the insertions $\drr$ and $\dll$ the effects of including beyond
leading order GFM contributions typically reduce the values of
BR($\bsm$) and $\delmbs$ compared to a purely LO calculation, 
exhibiting a focusing effect in the Higgs sector similar to the
one pointed out in the case of $\bsg$ in Ref.~\cite{OR2:bsg}.

In the second part of our analysis we have illustrated how these
effects can constrain the values of the flavour violating parameters
$\drl$ and $\drr$. The strong enhancement that supersymmetric
contributions to $\delmbs$ and BR($\bsm$) receive for non--zero $\drl$
and $\drr$, can lead to far stricter constraints
on these parameters than in an analysis that
just takes into account the effects that they have on $\bsg$.
\\
\\
\noindent
{\bf Acknowledgements}\\
\noindent We would like to thank D.~Demir, G.~Giudice and A.~Masiero
for helpful comments. J.F. has been supported by a PPARC Ph.D.
studentship and K.O. by a Korean government grant KRF PBRG 2002-070-C00022.

\begin{appendix}
\section{Loop Functions}\label{lfunc}
The loop functions $H_2(x_1,x_2)$ and $H_3(x_1,x_2,x_3)$ are given by
\begin{align}
H_2(x_1,x_2)=&\frac{x_1\log x_1}{\left(1-x_1\right)\left(x_1-x_2\right)}
+\frac{x_2\log x_2}{\left(1-x_2\right)\left(x_2-x_1\right)}\\
H_3(x_1,x_2,x_3)=&\frac{H_2(x_1,x_2)-H_2(x_1,x_3)}{x_2-x_3}
\end{align}
\end{appendix}


\begin{thebibliography} {99}
\bibitem{SLAC:bbm} SLAC Heavy Flavour Averaging Group
  http://www.slac.stanford.edu/xorg/hfag/.

\bibitem{GM:bsg} P.~Gambino and M.~Misiak,
  Nucl. Phys. {\bf B611}, 338 (2001) [hep-ph/0104034]. 

\bibitem{BCMU:bsg} A.~Buras, A.~Czarnecki, M.~Misiak and J.~Urban,
  Nucl. Phys. {\bf B631}, 219 (2002) [hep-ph/0203135].

\bibitem{HLP:bsg} T.~Hurth, E.~Lunghi and W.~Porod,
  hep-ph/0312260.

\bibitem{GHM:bsg} P.~Gambino, U.~Haisch and M.~Misiak,
  hep-ph/0410155.

\bibitem{LO:bsg} S.~Bertolini, F.~Borzumati, A.~Masiero and G.~Ridolfi,
  Nucl. Phys. {\bf B353}, 591 (1991);
  F.~Borzumati, C.~Greub, T.~Hurth and D.~Wyler,
  Phys. Rev. {\bf D62}, 075005 (2000) [hep-ph/9911245];
  T.~Besmer, C.~Greub and T.~Hurth,
  Nucl. Phys. {\bf B609}, 359 (2001) [hep-ph/0105292].

\bibitem{CDGG1:bsg} M.~Ciuchini, G.~Degrassi, P.~Gambino and G.~Giudice,
  {\sl Nucl. Phys.} {\bf B527}, 21 (1998) [hep-ph/9710335].

\bibitem{BG:bsg} F.~Borzumati and C.~Greub,
  Phys. Rev. {\bf D58}, 074004 (1998) [hep-ph/9802391].

\bibitem{CDGG2:bsg} M.~Ciuchini, G.~Degrassi, P.~Gambino and G.~Giudice,
  {\sl Nucl. Phys.} {\bf B534}, 3 (1998) [hep-ph/9806308].

\bibitem{DGG:bsg} G.~Degrassi, P.~Gambino and G.~Giudice,
  JHEP {\bf 0012}, 009 (2000) [hep-ph/0009337].

\bibitem{CGNW:bsg}
  M.~Carena, D.~Garcia, U.~Nierste and C.~E.~Wagner,
  Phys. Lett. {\bf B499}, 141 (2001) [hep-ph/0010003].

\bibitem{AGIS:bdec} G.~D'Ambrosio, G.~Giudice, G.~Isidori and A.~Strumia,
  Nucl. Phys. {\bf B645}, 155 (2002) [hep-ph/0207036].

\bibitem{BCRS:bdec} A.~Buras, P.~Chankowski, J.~Rosiek and
  {\L}.~S{\l}awianowska,
  Nucl. Phys. {\bf B659}, 3 (2003) [hep-ph/0210145].

\bibitem{OR2:bsg} K.~Okumura and L.~Roszkowski,
  Phys. Rev. Lett. {\bf 92}, 161801 (2004) [hep-ph/0208101] and
  JHEP {\bf 0310}, 024 (2003) [hep-ph/0308102]. 

\bibitem{D0:bsm} V.~M.~Abazov et al., D\O~Collaboration,
  hep-ex/0410039.

\bibitem{CDF:bsm} D.~Acosta et al., CDF Collaboration,
  Phys. Rev. Lett. {\bf 93} 032001, (2004) [hep-ex/0403032].

\bibitem{NLO:bsm} G.~Buchalla and A.~Buras,
  Nucl. Phys. {\bf B398}, 285 (1993),
  ibid. {\bf B400}, 225 (1993) and
  ibid. {\bf B548}, 309 (1999) [hep-ph/9901288];  
  M.~Misiak and J.~Urban,
  Phys. Lett. {\bf B451}, 161 (1999) [hep-ph/9901278].

\bibitem{NLO:bbm} A.~Buras, M.~Jamin and P.~Weisz,
  Nucl. Phys. {\bf B347}, 491 (1990). 

\bibitem{Buras:rev} A.~Buras, 
  hep-ph/0101336.
 
\bibitem{BCJL:bbm} V.~Barger, C-W. Chiang, J.~Jiang and P.~Langacker,
  Phys. Lett. {\bf B596}, 229 (2004) [hep-ph/0405108].

\bibitem{Buras:bsm} A.~Buras,
  Phys. Lett. {\bf B566}, 115 (2003) [hep-ph/0303060].

\bibitem{BK:bsm} K.~Babu and C.~Kolda,
  Phys. Rev. Lett. {\bf 84}, 228 (2000) [hep-ph/9909476]. 
 
\bibitem{MFV:bsm} C-S.~Huang, W.~Liao, Q-S.~Yan and S-H.~Zhu,
  Phys. Rev. {\bf D63}, 114021 (2001) [hep-ph/0006250];
  (E) {\sl ibid.} {\bf D64}, 059902 (2001);
  C.~Bobeth, T.~Ewerth, F.~Kr\"uger and J.~Urban,
  Phys. Rev. {\bf D64}, 074014 (2001) [hep-ph/0104284]. 

\bibitem{IR1:bsm} G.~Isidori and A.~Retico,
  JHEP {\bf 0111}, 001 (2001) [hep-ph/0110121].

\bibitem{GFM:bsm} P. Chankowski and {\L}. S{\l}awianowska,
  Phys. Rev. {\bf D63}, 054012 (2001) [hep-ph/0008046].

\bibitem{BEKU2:bsm} C.~Bobeth, T.~Ewerth, F.~Kr\"uger and J.~Urban,
  Phys. Rev. {\bf D66}, 074021 (2002) [hep-ph/0204225].

\bibitem{IR2:bsm} G.~Isidori and A.~Retico,
  JHEP {\bf 0209}, 063 (2002) [hep-ph/0208159].

\bibitem{DP:2002er} A.~Dedes and A.~Pilaftsis, 
  Phys. Rev. {\bf D67}, 015012 (2003) [hep-ph/0209306].  
 
\bibitem{Demir:bsm} D.~Demir,
  Phys. Lett. {\bf B571}, 193 (2003) [hep-ph/0303249].

\bibitem{BCRS:bbm} A.~Buras, P.~Chankowski, J.~Rosiek and
  {\L}.~S{\l}awianowska,
  Nucl. Phys. {\bf B619}, 434 (2001) [hep-ph/0107048]. 

\bibitem{BJU:bbm} A.~Buras, S.~J\"ager and J.~Urban,
  Nucl. Phys. {\bf B605}, 600 (2001) [hep-ph/0102316].

\bibitem{HRS:qme} L.~Hall, R.~Rattazzi and U.~Sarid,
  Phys. Rev. {\bf D50}, 7048 (1994) [hep-ph/9306309].

\bibitem{for2:bdec} J.~Foster, K.~Okumura and L.~Roszkowski, in
  preparation.

\bibitem{CFMS:bdec} M.~Ciuchini, E.~Franco, A.~Masiero and L.~Silvestrini,
  Phys. Rev. {\bf D67}, 075016 (2003) [hep-ph/0212397];
  (E) ibid. {\bf D68}, 079901 (2003).
  
\bibitem{CMM:neut} D.~Chang, A.~Masiero and H.~Murayama,
  Phys. Rev. {\bf D67}, 075013 (2003), [hep-ph/0205111].

\bibitem{drr:other}
  S.~Baek, T.~Goto, Y.~Okada and K.~Okumura, 
  Phys. Rev. {\bf D63}, 051701 (2001) [hep-ph/0002141]
  and 
  ibid. {\bf D64}, 095001 (2001) 
  [hep-ph/0104146];
%
  T.~Moroi,
  JHEP {\bf 0003}, 019 (2000) [hep-ph/0002208] 
  and 
  Phys. Lett. {\bf B493}, 366 (2000) [hep-ph/0007328].

\bibitem{FeynHiggs} M.~Frank, S.~Heinemeyer, W.~Hollik and G.~Weiglein,
  [hep-ph/0212037];
  S.~Heinemeyer,
  Eur. Phys. J. {\bf C22}, 521 (2001) [hep-ph/0108059];
  http://www.feynhiggs.de.

\bibitem{HHMP:pew} S.~Heinemeyer, W.~Hollik, F.~Merz and S.~Penaranda,
  hep-ph/0403228.

\end{thebibliography}
\end{document}